\documentclass[iop,apjl]{emulateapj}
\usepackage{amssymb}

\newcommand{\msun}{M$_\odot$}
\newcommand{\ha}{H$\alpha$}

\usepackage{color}

\shorttitle{A (nearly) uniform stellar initial mass function}
\shortauthors{Fumagalli et  al.}

\begin{document}


\title{Stochastic star formation and a (nearly) uniform stellar initial mass function}

\author{Michele Fumagalli \altaffilmark{1}, Robert L. da Silva  \altaffilmark{1,2}, 
and Mark R. Krumholz \altaffilmark{1}}
\email{mfumagalli@ucolick.org}


\altaffiltext{1}{Department of Astronomy and Astrophysics, University of California,  
1156 High Street, Santa Cruz, CA 95064.}
\altaffiltext{2}{NSF Graduate Research Fellow}


\begin{abstract}
Recent observations indicate a lower \ha\ to FUV ratio in
dwarf galaxies than in brighter systems, a trend that    
could be explained by a truncated and/or steeper IMF in small galaxies. 
However, at low star formation rates (SFRs), the \ha\ to FUV ratio can vary due to stochastic 
sampling even for a universal IMF, a hypothesis that has, prior to this work, 
received limited investigation. Using \textsc{slug}, a fully stochastic code for 
synthetic photometry in star clusters 
and galaxies, we compare the \ha\ and FUV luminosity in a sample of
$\sim450$ nearby galaxies with models drawn from a universal Kroupa IMF and
a modified IMF, the integrated galactic initial mass function (IGIMF). 
Once random sampling and time evolution are included, a Kroupa IMF convolved with the 
cluster mass function reproduces the observed $\rm H\alpha$ distribution 
at all FUV luminosities, while a truncated IMF as implemented in current IGIMF models 
underpredicts the $\rm H\alpha$ luminosity by more than an order of magnitude at 
the lowest SFRs. We conclude that the observed luminosity is the result of the 
joint probability distribution function of the SFR, cluster mass function, 
and a universal IMF, consistent with parts of the IGIMF theory, but that a 
truncation in the IMF in clusters is inconsistent with the observations.
Future work will examine stochastic star formation
and its time dependence  in detail to study whether random sampling can 
explain other observations that suggest a varying IMF. 
\end{abstract}


\section{Introduction}\label{intro}

The stellar initial mass function (IMF), an essential ingredient for numerous astrophysical
problems, is commonly assumed to be invariant with time and with galactic properties.
This \emph{ansatz} follows from a lack of evidence for its variation, 
despite searches covering a wide range of environments \citep[e.g.][]{kro01,cha03,bas10},
more than from theoretical understanding of the processes that regulate the 
mass distribution of stars. However, several recent studies have
questioned the idea of a universal IMF in both nearby and distant galaxies. 

Considering a few examples in the local universe, 
a truncated and/or steeper IMF has been invoked to explain 
an apparent correlation between the masses of clusters and the masses of their
largest members (\citealt{wei06}, though see \citealt{lam10} for an opposing view),  
or between galaxy colors and \ha\ equivalent widths  \citep{hov08,gun11},
as well as between the ratio of \ha\ to FUV luminosity ($L_{\rm H\alpha}/L_{\rm FUV}$)
and the surface brightness \citep{meu09} or the star formation rates \citep[SFRs;][]{lee09,wei11} 
in galaxies. However, these claims remain controversial, and
alternative interpretations have been offered, 
without resorting to IMF variations \citep{bos09,cor09,cal10,lam10}.

In this letter we focus on observations of systematic variations in $L_{\rm H\alpha}/L_{\rm FUV}$
\citep[e.g.][]{meu09,lee09,bos09} that have been used to argue strongly for
a non-universal IMF \citep[e.g.][]{meu09,wei11}. Recombination lines such as
\ha\ ultimately come from the ionizing photons produced primarily by very massive
stars, while FUV luminosity is driven by less massive stars ($\sim50$ \msun\
for \ha, $\sim10$ \msun\ for FUV, using a \citealt{sal55} IMF). As a result, the ratio
$L_{\rm H\alpha}/L_{\rm FUV}$ can be used as a probe for the massive end of the IMF, 
although the interpretation of observations is complicated by additional factors (e.g. dust).
Recently, both  \citet[][hereafter M09]{meu09} and \citet[][hereafter L09]{lee09} reported
that $L_{\rm H\alpha}/L_{\rm FUV}$ varies systematically with galaxy properties,
which may imply a variation of the IMF. \citet[][hereafter B09]{bos09} have
cautioned that dust correction or a bursty star formation history (SFH) can produce the
observed trends even if the underlying IMF is universal, but this idea has not been
explored in detail with models. While they cannot rule out variations in the IMF, 
L09 also argue that the underlying cause for the observed trend is not clear and 
stochastic effects at low SFR need to be explored further. 

Following these suggestions \citep[see also][]{wil97,cer03,haa10,eld10}, in this work we 
ask whether the observed $L_{\rm H\alpha}/L_{\rm FUV}$ is consistent 
with a random but incomplete sampling of a canonical or modified IMF. 
To test these hypotheses, we combine observations from B09, L09 and M09 with models from 
\textsc{slug}\footnote{http://sites.google.com/site/runslug/} 
\citep{fum10,das11}, a novel fully-stochastic code for 
synthetic photometry of stellar clusters and galaxies.

\section{The observed galaxy sample}\label{data}

We compile a sample of 457 galaxies from B09, L09
and M09 with integrated \ha\ and FUV luminosities, 
together with corrections for dust extinction and [\ion{N}{2}] contamination. 
Due to differential absorption, particular care is required 
when performing dust corrections. Here we follow the procedure adopted by each author in their 
studies, a choice motivated by the fact that the different dust extinctions
are consistent across the three samples. We refer to the original works for additional discussion. 

Similarly, a careful analysis of the selection biases should be a prerequisite 
for comparison with models. Unfortunately, these data-sets
have been assembled adopting different selection criteria (a nearly complete volume limited sample for L09,
a subset of galaxies selected from the \ion{H}{1} mass function for M09 and a sample 
of galaxies with multi-wavelength observations for B09) that are difficult to characterize. 
We will therefore emphasize only those results that are believed to be less sensitive to selection biases.

\begin{figure}
\centering
\plotone{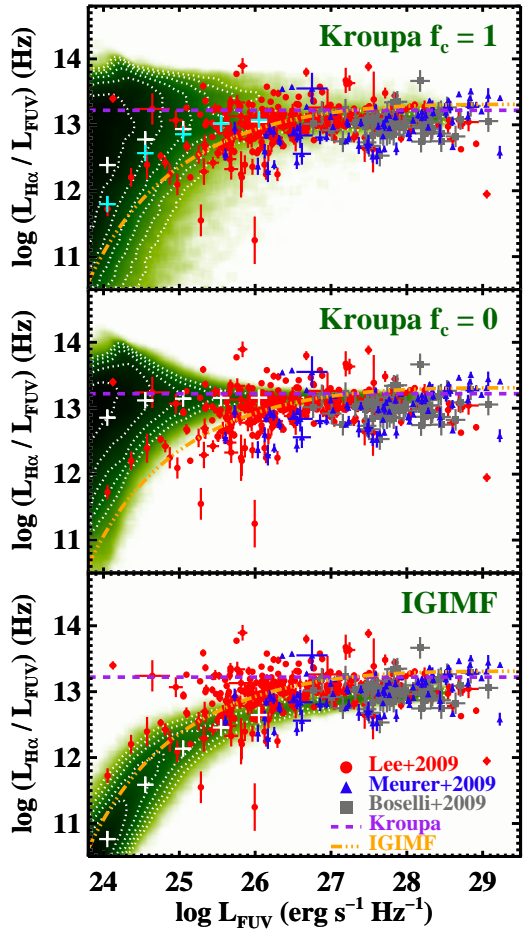}
\caption{Observed \ha\ and FUV luminosity in a sample of nearby galaxies (L09 red circles;
B09 gray squares; M09 blue triangles). We also show (in green), $\sim10^5$ \textsc{slug}
models for a Kroupa IMF with and without clusters (top and middle panel, respectively)
and for the IGIMF (bottom panel). Analytic predictions for the Kroupa IMF and IGIMF
are superimposed (purple dashed and orange triple-dot dashed lines). White crosses mark the 
mean of the simulated distributions, while the cyan crosses (top panel) are for a $f_{\rm c}=1$ model
with $M_{\rm cl,min}=500~\rm M_\odot$.}
\label{lumlum}
\end{figure}

\section{Stochastic effects and {\sc slug} models}\label{model}

The IMF describes the relative probability with which stars at any mass are formed. 
At high SFRs, the large number of stars guarantees a nearly complete sampling 
of the IMF, and the maximum stellar mass $m_{\rm max}$ that can be found in 
a galaxy approaches the theoretical upper end $m_{\rm max,*}$ of the IMF. 
At low SFRs instead, due to the smaller number of stars that are formed, 
the probability to find massive stars decreases. 
The IMF is not fully sampled, and $m_{\rm max}\ll m_{\rm max,*}$. In this regime, 
stochastic effects become important to describe the observed $L_{\rm H\alpha}/L_{\rm FUV}$ distribution, 
regardless of the IMF functional form. In addition to finite sampling in mass, $L_{\rm H\alpha}/L_{\rm FUV}$ is
also affected by finite sampling in time since massive stars experience evolutionary
phases of short duration (e.g.\ WN or WC phases) during which their ionizing luminosities
can vary significantly.

Since the majority of stars is believed to form in embedded star clusters \citep[e.g.][]{lad03},
two additional effects should be considered. First, the maximum stellar mass in a cluster 
cannot exceed the cluster mass $M_{\rm ecl}$. Second, at low SFRs, a time-averaged continuous SFR results in 
a series of small ``bursts" separated in time and associated with the formation of a new cluster. 
Provided that stars form with a modest spread in ages, gaps in 
the ages of clusters increase the probability 
to observe a galaxy when the most massive stars have already left the main sequence.

Because it treats these types of stochastic sampling correctly,
\textsc{slug} is the ideal tool to test whether incomplete sampling of a canonical IMF can reproduce the 
deficiency of \ha\ observed in low-luminosity galaxies without resorting to a modified IMF.
For any given SFH, a fraction $f_{\rm c}$ of the total stellar population  is assumed to form in clusters. 
In this case, {\sc slug} randomly draws a set of clusters from a cluster 
mass function (CMF) $\psi(M_{\rm ecl})=M_{\rm ecl}^{-\beta}$ and populates them 
with stars randomly selected from an IMF 
$\phi(m)=m^{-\gamma}$ over some specified mass interval. 
Currently, all the stars within a cluster are assumed to be coeval.
The remaining fraction $1-f_{\rm c}$ of the total stellar population are formed in the field,
simulated by randomly selecting stars from an IMF. The ensemble of clusters and stars is 
then evolved with time and at each time step the luminosity of individual stars is combined to 
quantify the number of ionizing photons and $L_{\rm FUV}$ for the simulated galaxy.

In this work, we consider two classes of models. The ``Kroupa model''
is based on a universal \citet{kro01} IMF in the mass 
interval $0.08-120$ M$_\odot$ and a CMF with $\beta=2$ between 
$20-10^7$ M$_\odot$. To test the effects of clustering we run two sets of simulations, one
with $f_{\rm c}=0$ (all stars in the field, similar to the Monte Carlo in L09) 
and one with $f_{\rm c}=1$ (all stars in clusters).
These extreme cases bracket all possible solutions, although the observations of \citet{lad03} 
suggest that reality is likely closer to $f_{\rm c}=1$ than $f_{\rm c}=0$.
The ``IGIMF model'' is instead based on the integrated galactic 
initial mass function \citep[IGIMF;][]{kro03}, 
a modified IMF in which $m_{\rm max}$ is a function of 
$M_{\rm ecl}$ that, in turn, depends on the galaxy SFR. 
The IGIMF implementation follows the minimal-1 and minimal-2 formulations \citep{pfl07},
with a \citet{kro01} IMF and a CMF with $\beta=2$, 
minimum mass of $20$ M$_\odot$ and maximum mass derived from the SFR averaged over $10^7$ year. 
In both the Kroupa $f_{\rm c}=1$ and the IGIMF model, the IMF integrated over a galaxy 
is modulated by the galaxy star formation and the CMF.
While in the $f_{\rm c}=1$ model the relations between these quantities emerge 
from a pure stochastic realization of the IMF and CMF, the IGIMF relies on prescriptions
for these correlations that imply a truncation in the CMF and the IMF.

For both models, we adopt stellar libraries with the Padova AGB tracks at solar metallicity
\citep{bre93}, the \citet{smi02} implementation of \citet{hil98} and \citet{lej97} atmospheres, 
and \citet{mae88} winds. We run $10^{5}$ simulated galaxies 
randomly drawn from a Schechter function, similar to the observed UV luminosity 
function \citep{wyd05}, with SFR $4.5\times10^{-5}-5~\rm  M_\odot\:yr^{-1}$. 
We use 10 time steps of 80 Myr per galaxy to improve the statistics.
Each of them is independent since the 80 Myr interval ensures that 
all massive stars that form in one time step are gone by the next one.
To further improve the statistics at higher luminosity (observations
are biased towards higher $L_{\rm FUV}$), we add 2000 galaxies uniformly 
distributed in logarithmic bins of SFR.
Due to escape or dust absorption before ionization, only a fraction 
$f_{\rm H\alpha}$ of the Lyman continuum photons produces ionizations that ultimately 
yield H$\alpha$ photons. To account for this effect, we correct the \ha\ luminosity 
in the models by a factor  $f_{\rm H\alpha}=0.95$ (B09).
Although the exact value for $f_{\rm H\alpha}$ is uncertain (L09), 
any choice in the range $f_{\rm H\alpha}=1-0.6$ \citep[e.g. B09,][]{hir03} would 
leave our basic results unchanged.
We further emphasize that reasonable choices of metallicity, stellar libraries, 
time steps, or even IMF (Kroupa versus Salpeter) yield very similar results.

\textsc{Slug} simulations and theoretical predictions are displayed in Figure \ref{lumlum}.
Stochastic effects are evident in the distributions, with simulated galaxies scattered in proximity to
the expected luminosities for a fully sampled IMF. 
Since previous studies \citep[e.g.][]{haa10} have highlighted 
the importance of the minimum mass in the CMF ($M_{\rm cl,min}$) for this type of 
calculations, we display in the top panel of Figure \ref{lumlum} the mean values for 
additional simulations that are similar to the $f_{\rm c}=1$ model, 
but with $M_{\rm cl,min}=500~\rm M_\odot$.

\section{Discussion}\label{discussion}

The main result of this analysis is summarized in Figure \ref{lumlum}. 
At $L_{\rm FUV}>10^{27} \rm ~erg\:s^{-1}\:Hz^{-1}$,  observed galaxies lie close to the value of
$L_{\rm H\alpha}/L_{\rm FUV}$ expected for both a fully-sampled Kroupa
IMF and the IGIMF, but at fainter luminosity and lower SFR
the data deviates from the fully-sampled Kroupa IMF 
curve towards the fully-sampled IGIMF curve. This trend has been 
taken as evidence in support of the IGIMF, but both IMFs are subject to 
stochastic effects. When we properly include these using \textsc{slug}, we see that
realizations drawn from the IGIMF are completely inconsistent with a significant 
fraction of the observed galaxies,
particularly at $L_{\rm FUV}\sim10^{25}-10^{26} \rm erg\:s^{-1}\:Hz^{-1}$.
Conversely, realizations drawn from a Kroupa IMF span a larger range of luminosity and 
overlap with most of the observed sample, with clustering responsible for a further increase
in the luminosity spread (compare the $f_{\rm c}=1$ and $f_{\rm c}=0$ models).

The width in the simulated distributions follows from 
the treatment of $m_{\rm max}$ and clustering. For a universal IMF, 
$m_{\rm max}$ can assume any value up to $m_{\rm max,*}$, regardless of the SFR. 
At low SFRs, realizations that lack massive stars
are frequent and skew the distribution to low $L_{\rm H\alpha}$ and low $L_{\rm FUV}$. At the same time, 
realizations with massive stars are still possible and some models are distributed 
near or even above the theoretical expectation for a fully sampled IMF.
The narrower scatter found for the $f_{\rm c}=0$ model emphasizes  
that stochastic sampling of the IMF alone cannot reproduce the entire range 
of observed luminosity (cf. L09). In the $f_{\rm c}=1$ model, simulated galaxies that lie at the
lowest $L_{\rm H\alpha}$ for any given $L_{\rm FUV}$ have an excess of older and massive clusters.
This is because at low SFR, a massive cluster represents a significant
event in the galaxy SFH that, on average, increases the time interval between the formations
of clusters and produces an intrinsic level of burstiness
that results in a wider luminosity distribution
\citep[see also][]{das11}. This effect is obviously 
amplified in models with $M_{\rm cl,min}=500~\rm M_\odot$, as evident from the lower 
H$\alpha/$FUV luminosity in these simulations. 
Note that irregular SFHs are typical of dwarf galaxies \citep{wei08} and 
models with bursty star formation may reproduce the observed luminosity distribution equally well.
Conversely, in the IGIMF theory, at lower SFRs, $m_{\rm max}\ll m_{\rm max,*}$. Further, only 
clusters with low mass can be drawn. The narrower mass range that is accessible translates in a narrower spread 
in luminosity, and none of the models can significantly exceed the expected luminosities for a fully 
sampled IGIMF (see L09).

\begin{deluxetable*}{ccccccc}
\tablecolumns{6}
\tablewidth{0pt}
\tablecaption{Summary of the \ha\ statistics in four bins of FUV luminosity from Figure \ref{histo}.}
\tablehead{
\colhead{Type\tablenotemark{a}}&
\colhead{Number\tablenotemark{b}}&
\colhead{Mean $\log L_{\rm H\alpha}$\tablenotemark{c}}&
\colhead{Fully-sampled $\log L_{\rm H\alpha}$\tablenotemark{d}}&
\colhead{Dispersion\tablenotemark{e}}&
\colhead{Probability\tablenotemark{f}}&
\colhead{N($\sigma$)\tablenotemark{g}}\\
\colhead{}&
\colhead{}&
\colhead{$\rm (erg~s^{-1})$}&
\colhead{$\rm (erg~s^{-1})$}&
\colhead{}&
\colhead{}&
\colhead{}} 
\startdata
\cutinhead{$24.0< \log L_{\rm FUV} <25.0$}
Data &           10 & 37.1 & \nodata  &  0.64 & \nodata &\nodata\\
Kroupa $f_{\rm c}=1$&       241817 & 37.0 & 37.7 &  0.86 & 0.5995& 0.5\\
Kroupa $f_{\rm c}=0$&       362593 & 37.3 & 37.7 &  0.55 & 0.5515& 0.6\\
IGIMF &        64253 & 35.6 & 36.4 &  0.72 & 0.0034& 2.9\\
\cutinhead{$25.0< \log L_{\rm FUV} <26.0$}
Data &           86 & 38.5 & \nodata  &  0.49 & \nodata &\nodata\\
Kroupa $f_{\rm c}=1$&        25571 & 38.3 & 38.7 &  0.55 & 0.1073& 1.6\\
Kroupa $f_{\rm c}=0$&        24705 & 38.5 & 38.7 &  0.30 & 0.1205& 1.6\\
IGIMF &         9363 & 37.7 & 38.2 &  0.45 & 0.0000& 6.9\\
\cutinhead{$26.0< \log L_{\rm FUV} <27.0$}
Data &          153 & 39.5 & \nodata  &  0.45 & \nodata &\nodata\\
Kroupa $f_{\rm c}=1$&         4248 & 39.5 & 39.7 &  0.40 & 0.3524& 0.9\\
Kroupa $f_{\rm c}=0$&         4238 & 39.6 & 39.7 &  0.29 & 0.0002& 3.7\\
IGIMF &         3576 & 39.2 & 39.6 &  0.37 & 0.0003& 3.6\\
\cutinhead{$27.0< \log L_{\rm FUV} <28.0$}
Data &          135 & 40.5 & \nodata  &  0.35 & \nodata &\nodata\\
Kroupa $f_{\rm c}=1$&         3494 & 40.6 & 40.7 &  0.34 & 0.4042& 0.8\\
Kroupa $f_{\rm c}=0$&         3562 & 40.6 & 40.7 &  0.29 & 0.0090& 2.6\\
IGIMF &         3296 & 40.4 & 40.7 &  0.33 & 0.0365& 2.1\\
\enddata\label{tab1}
\tablenotetext{a}{Type of distribution for which the statistics are listed.}
\tablenotetext{b}{Number of observed and simulated galaxies included in each $\log L_{\rm FUV}$ bin.}
\tablenotetext{c}{Mean of the $\log L_{\rm H\alpha}$ distributions.}
\tablenotetext{d}{$\log L_{\rm H\alpha}$ for a fully sampled IMF.}
\tablenotetext{e}{Standard deviation of the $\log L_{\rm H\alpha}$ distributions.}
\tablenotetext{f}{Kolmogorov-Smirnov probability $P$ associated to the hypothesis that 
the observed and simulated distributions are drawn from the same parent population.
Although the absolute values for the listed probabilities are difficult to interpret 
given the poorly characterized selection biases, the relative differences between 
the $f_{\rm c}=1$ and the IGIMF models support quantitatively our conclusion.}
\tablenotetext{g}{Equivalent number of standard deviations $N=\sqrt2\rm~erfc^{-1}(P)$.}
\end{deluxetable*}

\begin{figure}
\centering
\plotone{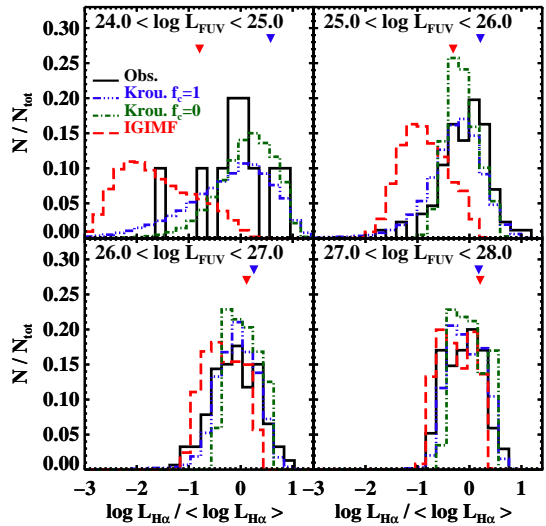}
\caption{Distributions of \ha\ luminosity in intervals of $L_{\rm FUV}$
for the observed galaxies (solid black histogram) and for {\sc slug} models based on the 
Kroupa IMF with and without clusters (blue dash-triple-dotted and green dash-dotted histograms) and for 
the IGIMF (red dashed histogram). Distributions are centered to $\langle L_{\rm H\alpha}\rangle$, the 
logarithmic mean of the observed \ha\ luminosity in each bin. 
The two downward triangles indicate the mean $L_{\rm H\alpha}$ for a 
fully sampled Kroupa IMF (blue) and IGIMF (red).}
\label{histo}
\end{figure}

Due to incompleteness, statistical comparisons between data and simulations 
are not straightforward. This is complicated by the fact that, at any SFR,
both \ha\ and FUV luminosities are subject to scatter and a truly independent variable 
is lacking. Nevertheless, in Figure \ref{histo} and Table \ref{tab1},
we attempt to quantify the agreement between models and 
observations by comparing the statistics of the \ha\ distributions in intervals of $L_{\rm FUV}$. 

Considering the center of the \ha\ distributions, the IGIMF and 
$f_{\rm c}=1$ realizations diverge, moving towards lower FUV luminosity. At the lowest $L_{\rm FUV}$, 
the two distributions are separated by $\sim2$ dex, with the data clearly favoring the Kroupa IMF.
Figure \ref{histo} demonstrates that a truncated IMF and CMF
as in the IGIMF produce systematically lower $L_{\rm H\alpha}$ than observed.
This is particularly evident for the interval $25<\log L_{\rm FUV}<26$
that is characterized by large enough statistics  ($\sim100$ galaxies) and 
likely not affected by severe incompleteness. To quantify this claim, 
we perform a Kolmogorov-Smirnov test comparing the observed distribution 
with 10000 random sub-samples of the models, 
extracted to have a size comparable to the data.
We report the results in Table \ref{tab1}. We see that the observed
data and the Kroupa $f_{\rm c}=1$ models are generally
consistent with being drawn from the same parent distribution, while
the hypothesis that the data and the IGIMF models originate from the
same parent distribution can be ruled out. 
The $f_{\rm c}=0$ model also appears to be inconsistent with the data, 
though not by as much as the IGIMF model.

\begin{figure}
\centering
\plotone{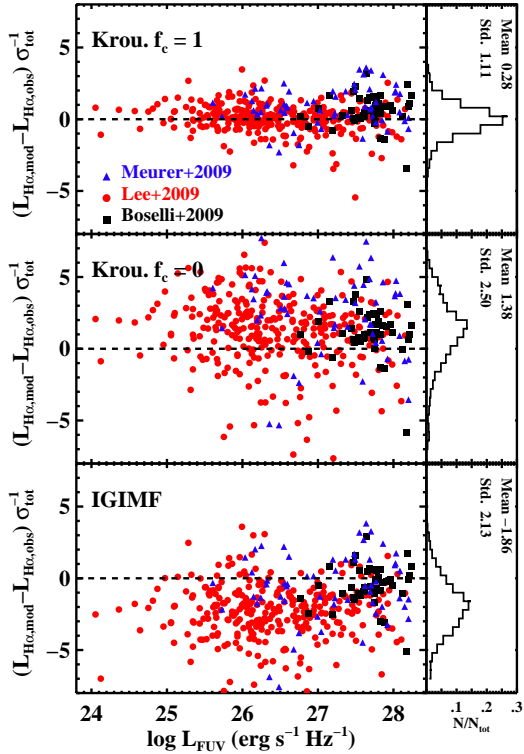}
\caption{Difference between $L_{\rm H\alpha}$ in individual observations
and the mean of the simulated galaxies at comparable $L_{\rm FUV}$,
normalized to the model standard deviation and observational uncertainty. 
Histograms and Gaussian statistics are shown in the right panels.}
\label{chi}
\end{figure}

Finally, in Figure \ref{chi} we compare for each individual data point 
the observed $L_{\rm H\alpha,obs}$ with the simulated $L_{\rm H\alpha,mod}$,
averaged over an interval of FUV that is the larger of 0.1 dex and the error on
the observed $L_{\rm FUV}$. This difference is then normalized to 
$\sigma_{\rm tot}=\sqrt{\sigma_{\rm obs}^2+\sigma_{\rm mod}^2}$
where $\sigma_{\rm obs}$ is the error on $L_{\rm H\alpha,obs}$ 
and $\sigma_{\rm mod}$ is the standard deviation of the models. 
This quantity is defined such that a distribution 
centered on zero with dispersion of unity indicates perfect agreement 
between models and data (see right panels).

In Figure \ref{chi}, the center of the  $f_{\rm c}=1$ model (top panel)
is consistent with the observed galaxies within 2$\sigma_{\rm tot}$ 
for $L_{\rm FUV}<10^{27} \rm ~erg\:s^{-1}\:Hz^{-1}$ and 
within 3$\sigma_{\rm tot}$ for brighter luminosity, where $\sigma_{\rm obs}$ becomes comparable to $\sigma_{\rm mod}$.
Conversely, due to the narrower scatter, the  $f_{\rm c}=0$ model (central panel) 
is only partially consistent with observations, as previously suggested by the KS test. 
Also, simulated galaxies have higher $L_{\rm H\alpha}$ compared to the observed galaxies. 
Based on results of two independent tests, we favor the $f_{\rm c}=1$ model over the 
$f_{\rm c}=0$ simulations. The IGIMF model (bottom panel) reproduces instead  
the observed $L_{\rm H\alpha}$ only at high $L_{\rm FUV}$
and there is a clear systematic offset between the data and the models 
at $L_{\rm FUV}<10^{27} \rm ~erg\:s^{-1}\:Hz^{-1}$. Although the usual notion of probability 
associated to $\sigma_{\rm tot}$ does not apply since the \ha\ distributions are not Gaussian, 
we conclude that observations are better described by models based on the Kroupa IMF 
than on the IGIMF, particularly at fainter luminosity and when clusters are included.

Data are very sparse below $L_{\rm FUV}\sim10^{25} \rm ~erg\:s^{-1}\:Hz^{-1}$, 
where the separation between a universal IMF and a steeper/truncated IMF  
is most evident, precluding us from concluding that no variation in the IMF occurs at 
very low luminosity. Moreover, we have not explored correlations with the
galaxy surface brightness, suggested by M09 to be the physical quantity related to the IMF variation.
Also, selection biases are not well characterized and a putative population of 
galaxies with $L_{\rm H\alpha}<10^{36} \rm ~erg\:s^{-1}$ and 
$L_{\rm FUV}\sim10^{25}-10^{26} \rm ~erg\:s^{-1}\:Hz^{-1}$ would call into question a universal IMF. 
Further we assume that all stars born in an individual cluster are coeval, and future work will be 
required to investigate the effects of relaxing this assumption.
Conversely, the observed existence of galaxies with $L_{\rm H\alpha}\sim10^{38}-10^{39} \rm ~erg\:s^{-1}$ 
and $L_{\rm FUV}\sim10^{24}-10^{26} \rm ~erg\:s^{-1}\:Hz^{-1}$ poses a direct challenge to 
the truncation in the IMF and CMF as currently implemented in IGIMF theory. 
As illustrated in Figure \ref{lumlum}, if the IGIMF model is correct,
it should be impossible for galaxies to occupy this region of luminosity space.
The conclusion is strengthened by the fact that we have implemented only a minimal IGIMF.
The standard IGIMF would predict even less $L_{\rm H\alpha}$, exacerbating the discrepancy with the observations.
Uncertainties in the measurements, particularly in the dust corrections,
together with the fact that our models are highly idealized (simple SFH, single metallicity, 
lack of any feedback), may explain the existence of galaxies non-overlapping with IGIMF models,
but the discrepancy we found is most likely larger than the errors at faint $L_{\rm H\alpha}$.

\section{Summary and conclusion}\label{conclu}

Using \textsc{slug}, a fully stochastic code for synthetic photometry in star clusters 
and galaxies, we have compared the \ha\ to FUV luminosities
in a sample of $\sim450$ nearby galaxies with models from a universal Kroupa IMF and
a modified IMF, the IGIMF. Our principal findings are: i) simulated galaxies 
based on a Kroupa IMF and stochastic sampling of stellar masses 
are consistent with the observed $L_{\rm H\alpha}$ distribution;
ii) only models where stars are formed in clusters account for the 
full scatter in the observed luminosity; iii) realizations based on a truncated IMF as currently 
implemented in the IGIMF underestimate the mean \ha\ luminosity.

Based on this result, and since other factors not included in our simulations
(e.g. dust, escape fraction of ionizing 
radiation, bursty SFHs) can mimic some of the features of the  
observed luminosity distribution (B09, L09), we conclude that present observations 
of the integrated luminosity in nearby galaxies are consistent with a universal IMF 
and do not demand a truncation at its upper end.

While we show that the current IGIMF implementation provides a poor description 
of available observations, our work is consistent with the 
fundamental idea behind the IGIMF, i.e.\ that the SFR,  the CMF, and the 
IMF jointly produce the observed luminosity distribution of galaxies. 
However, our analysis emphasizes that the correlations between these quantities emerge 
naturally from stochastic sampling and that a further modification to the IMF in 
clusters as proposed in the IGIMF model is not needed to account  for the integrated 
luminosities in galaxies. Further, our calculation highlights 
how clusters introduce the level of burstiness required to fully account for the observed 
luminosities, owing to the combined effects of the cluster age distribution 
and the short life-time of massive stars. Time dependence might be a 
key element currently missing in models of the integrated IMF in galaxies.

\acknowledgments
We are most grateful to A.Boselli and G.Meurer for sharing their data
and for valuable comments. We acknowledge useful discussions with 
M.Cervi\~no, J.Lee, P.Kroupa, J.Scalo, X.Prochaska, J.Werk, G.Gavazzi and K.Schlaufman.
We thank F.Bigiel for motivating us to write \textsc{slug}.
RLdS is supported under a NSF Graduate Research Fellowship.
MRK acknowledges support from: Alfred P.\ Sloan Fellowship; NSF grants
AST-0807739 and CAREER-0955300; NASA Astrophysics Theory and 
Fundamental Physics grant NNX09AK31G; {\it Spitzer Space Telescope} Theoretical
Research Program grant.


\begin{thebibliography}{}
\bibitem[Bastian et al.(2010)]{bas10} Bastian, N., Covey, K.~R., \& Meyer, M.~R.\ 2010, \araa, 48, 339
\bibitem[Boselli et al.(2009)]{bos09} Boselli, A., Boissier, S., Cortese, L., Buat, V., Hughes, T.~M., \& Gavazzi, G.\ 2009, \apj, 706, 1527 
\bibitem[Bressan et al.(1993)]{bre93} Bressan, A., Fagotto, F., Bertelli, G., \& Chiosi, C.\ 1993, \aaps, 100, 647 
\bibitem[Calzetti et al.(2010)]{cal10} Calzetti, D., Chandar, R., Lee, J.~C., Elmegreen, B.~G., Kennicutt, R.~C., \& Whitmore, B.\ 2010, \apjl, 719, L158 
\bibitem[Cervi{\~n}o \& Valls-Gabaud(2003)]{cer03} Cervi{\~n}o, M., \& Valls-Gabaud, D.\ 2003, \mnras, 338, 481 
\bibitem[Chabrier(2003)]{cha03} Chabrier, G.\ 2003, \pasp, 115, 763
\bibitem[Corbelli et al.(2009)]{cor09} Corbelli, E., Verley, S., Elmegreen, B.~G., \& Giovanardi, C.\ 2009, \aap, 495, 479 
\bibitem[da Silva et al.(2011)]{das11} da Silva, R.~L., Fumagalli, M., \& Krumholz, M.\ 2011, arXiv:1106.3072 
\bibitem[Eldridge(2011)]{eld10} Eldridge, J.~J.\ 2011, arXiv:1106.4311 
\bibitem[Fumagalli et al.(2011)]{fum10} Fumagalli, M., da 
Silva, R., Krumholz, M., \& Bigiel, F.\ 2011, Astronomical Society of the Pacific Conference Series, 440, 155 
\bibitem[Gunawardhana et al.(2011)]{gun11} Gunawardhana, M.~L., et al.\ 2011, \mnras, in press, arXiv:1104.2379
\bibitem[Haas \& Anders(2010)]{haa10} Haas, M.~R., \& Anders, P.\ 2010, \aap, 512, A79 
\bibitem[Hillier \& Miller(1998)]{hil98} Hillier, D.~J., \& Miller, D.~L.\ 1998, \apj, 496, 407 
\bibitem[Hirashita et al.(2003)]{hir03} Hirashita, H., Buat, V., \& Inoue, A.~K.\ 2003, \aap, 410, 83 
\bibitem[Hoversten \& Glazebrook(2008)]{hov08} Hoversten, E.~A., \& Glazebrook, K.\ 2008, \apj, 675, 163 
\bibitem[Kennicutt(1998)]{ken98} Kennicutt, R.~C., Jr.\ 1998, \araa, 36, 189 
\bibitem[Kroupa(2001)]{kro01} Kroupa, P.\ 2001, \mnras, 322, 231 
\bibitem[Kroupa \& Weidner(2003)]{kro03} Kroupa, P., \& Weidner, C.\ 2003, \apj, 598, 1076 
\bibitem[Lada \& Lada(2003)]{lad03} Lada, C.~J., \& Lada, E.~A.\ 2003, \araa, 41, 57 
\bibitem[Lamb et al.(2010)]{lam10} Lamb, J.~B., Oey, M.~S., Werk, J.~K., \& Ingleby, L.~D.\ 2010, \apj, 725, 1886 
\bibitem[Lee et al.(2009)]{lee09} Lee, J.~C., et al.\ 2009, \apj, 706, 599 
\bibitem[Lee et al.(2010)]{lee10} Lee, J.~C., Gil de Paz, A., Tremonti, C., Kennicutt, R., \& the Local Volume Legacy Team 2010, arXiv:1011.2181 
\bibitem[Lejeune et al.(1997)]{lej97} Lejeune, T., Cuisinier, F., \& Buser, R.\ 1997, \aaps, 125, 229 
\bibitem[Maeder \& Meynet(1988)]{mae88} Maeder, A., \& Meynet, G.\ 1988, \aaps, 76, 411 
\bibitem[Meurer et al.(2009)]{meu09} Meurer, G.~R., et al.\ 2009, \apj, 695, 765 
\bibitem[Pflamm-Altenburg et al.(2007)]{pfl07} Pflamm-Altenburg, J., Weidner, C., \& Kroupa, P.\ 2007, \apj, 671, 1550 
\bibitem[Pflamm-Altenburg et al.(2009)]{pfl09} Pflamm-Altenburg, J., Weidner, C., \& Kroupa, P.\ 2009, \mnras, 395, 394 
\bibitem[Salpeter(1955)]{sal55} Salpeter, E.~E.\ 1955, \apj, 121, 161 
\bibitem[Smith et al.(2002)]{smi02} Smith, L.~J., Norris, R.~P.~F., \& Crowther, P.~A.\ 2002, \mnras, 337, 1309 
\bibitem[Weidner \& Kroupa(2006)]{wei06} Weidner, C., \& Kroupa, P.\ 2006, \mnras, 365, 1333 
\bibitem[Weidner et al.(2011)]{wei11} Weidner, C., Kroupa, P., \& Pflamm-Altenburg, J.\ 2011, \mnras, 412, 979 
\bibitem[Weisz et al.(2008)]{wei08} Weisz, D.~R., Skillman, E.~D., Cannon, J.~M., et al.\ 2008, \apj, 689, 160 
\bibitem[Williams \& McKee(1997)]{wil97} Williams, J.~P., \& McKee, C.~F.\ 1997, \apj, 476, 166 
\bibitem[Wyder et al.(2005)]{wyd05} Wyder, T.~K., et al.\ 2005, \apjl, 619, L15 

\end{thebibliography}
\end{document}